\documentclass[aps,twocolumn]{revtex4}

\usepackage{graphicx}
\usepackage{graphics}
\usepackage{epsfig}

\begin{document}

\title{Time evolution of entangled biatomic states in a cavity}
\author{E. G. Figueiredo$^{a,d}$, C. A.
Linhares$^b$, A. P. C. Malbouisson$^c$ and J. M. C. Malbouisson$^a$}
\date{\today }

\begin{abstract}

We study the time evolution of entangled states of a pair of
identical atoms, considered in the harmonic approximation, coupled
to an environment represented by an infinite set of free
oscillators, with the whole system confined within a spherical
cavity of radius $R$. Taking the center-of-mass and the
relative-position coordinates, and using the dressed-state approach,
we present the time evolution of some quantities measuring the
entanglement, for both limits of a very large and a small cavity;
the chosen examples are simple and illustrate these very distinct
behaviors.

\noindent PACS number(s): 37.30.+i; 03.67.Mn; 42.50.Pq; 03.65.Ud
\end{abstract}

\affiliation{$^a$Instituto de F\'{\i}sica, Universidade Federal da
Bahia, 40210-340, Salvador, BA, Brazil\\$^b$Instituto de
F\'{\i}sica, Universidade do Estado do Rio de Janeiro, 20559-900,
Rio de Janeiro, RJ, Brazil\\$^c$Centro Brasileiro de Pesquisas
F\'{\i}sicas/MCT, 22290-180, Rio de Janeiro, RJ, Brazil\\$^d$Centro
de Ci\^encias Exatas e Tecnol\'ogicas, Universidade Federal do
Rec\^oncavo da Bahia, 44380-000, Cruz das Almas, BA, Brazil}
\maketitle

{\it Introduction}

Interacting or noninteracting parts of a quantum system can share
entangled states that hold quantum correlations~\cite{bel1,bel2}. An
recent review covering all relevant aspects of the subject is
in~\cite{horodecki}. Entanglement is a purely quantum phenomenon due
to the attribution of physical meaning to superposed states, a
concept with no correspondence in classical physics. Entanglement
means that individual parts of a quantum system are not independent
of each other, even if they do not interact, and their quantum
properties are described by their common wavefunction. In
particular, entanglement properties of bipartite systems have been
largely investigated along recent years. Within this context,  we
consider here a simple biatomic system, in which each atom is
modeled by a harmonic oscillator. In this case, studies have been
performed for a noninteracting bipartite system, with different
approaches, as for instance in Refs.~\cite{emaranho,outro cara}  and
when an interaction between the oscillators is taken into
account~\cite{pazes1,pazes2,prausner,dodd,benatti,ban,liu,an,horhammer}.

In this Brief Report, we study the time evolution of a superposition
of two biatomic states of identical atoms which interact indirectly
via the coupling with the harmonic modes of a field force
representing the environment. We consider the two atoms in the
harmonic approximation and assume that the whole system resides
inside a spherical cavity of radius $R$. Our basic objects will be
\textit{dressed} states, corresponding to the atoms dressed by the
field. The biatomic system will be consistently described by the
pair consisting of the ``center-of-mass"  and the
``relative-position" oscillators, a procedure already employed in
the literature, as for instance in Refs.~\cite{pazes1,pazes2}. In
our case, these oscillators will be appropriately dressed by the
field. We will consider the entangled state formed by the
superposition of two kinds of states: one state in which the
``center-of-mass" oscillator is in its first excited level and the
``relative-position" oscillator is in the ground state; this state
is  superposed with another state in which the oscillators have
their role reversed. We will be concerned by the system at zero
temperature, i.e. all the field modes are in their ground states.
Actually, it has been  shown in~\cite{addendum,termica1} that
thermal effects on dressed oscillators are important only for high
temperatures, they are negligible for room temperatures. This means
that we can make the approximation of taking the time evolution of
dressed states, as established for zero-temperature~\cite{emaranho}.
Thus, our results apply in fact for temperatures up to the order of
room temperatures.

{\it The model}

We start from a slightly modified version of the Hamiltonian used in
Refs.~\cite{pazes1,pazes2}, describing two atoms $A$ and $B$, in the
harmonic approximation, coupled to an environment, the whole system
being contained in a perfectly reflecting cavity of radius
$R$~\cite{footnote},
\begin{eqnarray}
H&=&\frac 12\left[ p_A^2+\omega _A^2q_A^2+p_B^2 + \omega _B^2q_B^2 +
\sum_{k=1}^N \left( p_k^2 + \omega _k^2q_k^2\right) \right]  \nonumber \\
&&-\sum_{k=1}^N\sqrt{2}(\eta _Aq_A+\eta _Bq_B)\omega _kq_k,
\label{Hamiltoniana}
\end{eqnarray}
where the limit $N\rightarrow \infty $ will be understood. Since the
atoms are identical, we write $ \omega_A = \omega _B \equiv \omega
_0$ and  $\eta _A = \eta _B \equiv \eta  = 2\sqrt{g\Delta \omega
/\pi }$, where $g$ is a constant with dimension of frequency
measuring the strength of the coupling and $\Delta \omega =\pi c/R$
is the interval between two neighboring frequencies of the
field~\cite{adolfo1}.

For identical atoms, we define new coordinates, $q_{+}$ (center of
mass) and $q_{-}$ (relative position), such that
\begin{equation}
q_{+} = \frac 1{\sqrt{2}}(q_A+q_B)\,;\;\;q_{-}=\frac 1{\sqrt{2}}(q_A-q_B),
\label{q+-}
\end{equation}
and corresponding formulas for momenta. Then in terms of $p_{\pm }$,
$q_{\pm }$, the Hamiltonian is written as $H = H_{-} + H_{+},$ where
$H_{-} = \frac 12\left[ p_{-}^2 + \omega _0^2q_{-}^2\right]$ and
\begin{equation}
H_{+} = \frac 12\left[ p_{+}^2+\omega _0^2q_{+}^2+\sum_{k=1}^N\left(
p_k^2+\omega _k^2q_k^2-\eta q_{+}\omega _kq_k\right) \right] .
\label{Hamiltoniana2}
\end{equation}
We see that the ``center-of-mass'' oscillator, $q_{+}$, couples to
the field while the ``relative-position" one, $q_{-}$, oscillates
freely. The Hamiltonian $H_{+}$ describes a single oscillator
linearly coupled to the field. Therefore, for the system $(q_{+}
\oplus field)$, composed by the ``center-of-mass'' oscillator
coupled to the field, we can define the renormalized frequency
$\bar{\omega}$, dressed coordinates and dressed states. Dressed
coordinates and states are defined by transformations of the
normal-mode coordinates and of the eigenstates of the diagonalized
Hamiltonian, which depend on the normal-mode frequencies and on the
renormalized oscillator frequency. An important aspect is that
excited dressed states decay with time while normal modes are
stationary. Details of this formalism are presented in
Refs.~\cite{emaranho,termica1,termica,adolfo1} and references
therein.

{\it Time evolution of entangled biatomic states}

We shall describe the pair of atoms by the pair center-of-mass and
relative-position oscillators. Consider the product states $| \Gamma
_{10}^{(+-)} \rangle \equiv | \Gamma _1^{(+)} \rangle \otimes |
\Gamma _0^{(-)} \rangle$ and $| \Gamma _{01}^{(+-)}\rangle \equiv |
\Gamma _0^{(+)}\rangle \otimes | \Gamma _1^{(-)}\rangle$ in which,
the dressed center-of-mass oscillator is in the first excited level,
while the relative-position oscillator is in its ground state, and
vice-versa, respectively. The dressed first-excited state of the
center-of-mass oscillator evolves in time according
to~\cite{adolfo1,emaranho}
\begin{equation}
| \Gamma _1^{(+)}(t) \rangle = \sum_\nu f_{+\nu }(t)| \Gamma _1^\nu
(0) \rangle , \label{ortovestidas54}
\end{equation}
with $\sum_\nu \left| f_{+ \nu }(t)\right| ^2 = 1$, where the label
$\nu$ runs over the center-of-mass oscillator $(+)$ and the field
modes $(\{ k \})$. The quantity $f_{+ \nu }(t)$ thus represents the
probability amplitude that the excitation is at the $\nu$-th dressed
oscillator at time $t$. On the other hand, the first excited state
of the relative-position oscillator is stationary. This implies that
the time evolution of the state $| \Gamma _{10}^{(+-)} \rangle $ is
governed by Eq.~(\ref{ortovestidas54}), while the state $| \Gamma
_{01}^{(+-)}\rangle $ remains stationary.

We now consider at $t=0$ the family of states
\begin{equation}
| \Psi ^{AB}(0) \rangle = \sqrt{\xi }\,| \Gamma _{10}^{(+-)}(0)
\rangle + \sqrt{1-\xi }\,e^{i\phi } | \Gamma _{01}^{(+-)}(0) \rangle
 \label{entangled}
\end{equation}
which belongs to the Hilbert space $\mathcal{H}_{+,\{k\}}\otimes
\mathcal{H}_{-}$, representing states of the system of the two atoms
coupled to the environmental field. At the instant $t$, the density
matrix corresponding to Eq~(\ref{entangled}) is written as
\begin{eqnarray}
\varrho (t) &=&| \Psi ^{AB}(t) \rangle \langle \Psi ^{AB}(t) | = \xi
 | \Gamma _{10}^{(+-)}(t) \rangle  \langle \Gamma
_{10}^{(+-)}(t) |  \nonumber \\
&&+\, ( 1-\xi  )  | \Gamma _{01}^{(+-)} \rangle
 \langle \Gamma _{01}^{(+-)} |  \nonumber \\
&&+\,\sqrt{\xi (1-\xi )}\,e^{i\phi } | \Gamma _{01}^{(+-)} \rangle
 \langle \Gamma _{10}^{(+-)}(t) |  \nonumber \\
&&+\,\sqrt{\xi (1-\xi )}\,e^{-i\phi } | \Gamma _{10}^{(+-)}(t)
\rangle \langle \Gamma _{01}^{(+-)} | . \label{densidade}
\end{eqnarray}
We adopt a more explicit notation and write $ | \Gamma
_{10}^{(+-)}(t) \rangle \equiv  | 1_{+}(t),0_{-};0,0,\ldots
 \rangle $ and $ | \Gamma _{01}^{(+-)} \rangle \equiv  |
0_{+},1_{-};0,0,\ldots  \rangle$. To analyze how the two-atom state
evolves in time, we consider the reduced density matrix, obtained by
taking the trace over the field modes,  $\rho(t) = \sum_{k_i}\langle
k_1,k_2,\ldots |\varrho (t)|k_1,k_2,\ldots \rangle $. Using
Eq.~(\ref{ortovestidas54}),
\begin{eqnarray*}
| 1_{+}(t),0_{-};0,0,\ldots \rangle = f_{++}(t) |
1_{+}(0),0_{-};0,0,\ldots \rangle \;\;\;\;\; \\
+\sum_if_{+i}(t) | 0_{+},0_{-};0,\ldots,0,1_i,0,\ldots \rangle ,
\end{eqnarray*}
and we obtain
\begin{equation}
{\rho }(t)=\left(
\begin{array}{cccc}
a(t) & 0 & 0 & 0 \\
0 & b(t) & d(t) & 0 \\
0 & d^{*}(t) & c(t) & 0 \\
0 & 0 & 0 & 0
\end{array}
\right) ,  \label{rhored}
\end{equation}
where \begin{eqnarray}
a(t) &\equiv &\left( \rho \right) _{0_{+}0_{-}}^{0_{+}0_{-}}=\xi \left(
1-\left| f_{++}(t)\right| ^2\right) \\
b(t) &\equiv &\left( \rho \right) _{0_{+}1_{-}}^{0_{+}1_{-}}=1-\xi ;
\label{elementos} \\
c(t) &\equiv &\left( \rho \right) _{1_{+}0_{-}}^{1_{+}0_{-}}=\xi \left|
f_{++}(t)\right| ^2; \\
d(t) &\equiv &\left( \rho \right) _{0_{+}1_{-}}^{1_{+}0_{-}}=\sqrt{\xi
(1-\xi )}e^{i\phi }f_{++}^{*}(t); \\
d^{*}(t) &\equiv &\left( \rho \right) _{1_{+}0_{-}}^{0_{+}1_{-}}=\sqrt{\xi
(1-\xi )}e^{-i\phi }f_{++}(t).
\end{eqnarray}
As it should, $\mathrm{Tr}\rho (t)=a(t)+b(t)+c(t)=1$. The degree of
impurity of the state (\ref{rhored}) is given by $D = 1 -
\mathrm{Tr}\rho ^2(t)$; then, it follows that
\begin{equation}
D(\xi ;t) = 2\xi ( 1-| f_{++}(t)| ^2)( 1-\xi + \xi | f_{++}(t)| ^2)
. \label{degrau1}
\end{equation}

To find the time dependence of the above quantities, we need to
evaluate the function $f_{++}(t)$ which governs the behavior of the
system. There are two significantly different situations, depending
on the size of the cavity. We shall analyze the limit of a very
large cavity ($R\rightarrow \infty$) and the case of a small cavity.
In any case, independently of emission frequency, we must have
$0<|f_{++}(t)|^2\leq 1$.

\noindent{\it Large cavity}: for a very large cavity, we
have~\cite{linhares}
\begin{equation}
f_{++}(t) = e^{-gt}\left[ \cos{\kappa t} - \frac{g}{\kappa} \sin{\kappa t}
\right] + i G(t;\bar{\omega},g) , \label{eq27}
\end{equation}
where $\kappa ^2=\bar{\omega}^2-g^2$ and the function $G$ is given
by
\begin{equation}
G(t;\bar{\omega},g) = - \frac{4g}{\pi}\int_0^\infty
dy\frac{y^2\sin{yt}}{(y^2 - \bar{\omega}^2)^2 + 4 g^2 y^2} .
\label{J}
\end{equation}
We will consider $\kappa^{2}>0$, which includes the weak coupling
regime $g^{2} \ll \bar{\omega}^2$.

\begin{figure}[th]
\includegraphics[{height=5.5cm,width=8.0cm}]{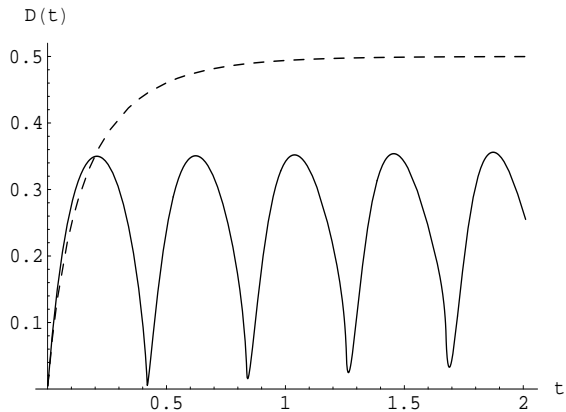}
\caption{Degree of impurity as function of time,
Eq.~(\ref{degrau1}), taking ${\bar \omega} = 1.5$ and $g = 1.0$ (in
arbitrary units), for states with $\xi = 0.5$: for a very large
cavity (dashed line) and for a small cavity with $\delta = 0.1$
(full line).} \label{impu}
\end{figure}

\noindent{\it Small cavity}: for a finite cavity, the spectrum of
eigenfrequencies is discrete, and the continuum language used in the
case of large cavity is not valid. However, for a very small cavity,
with a radius $R$ much smaller than the coherence length $\pi c/g$
($R\ll \pi c/g$), we can obtain $f_{++}(t)$ to a good approximation
as~\cite{linhares}
\begin{eqnarray}
f_{++}(t) & \approx & \left( 1+\frac 23\pi \delta \right)^{-1} \left\{ \exp
\left[ - i \bar{\omega}\left( 1 - \frac{\pi\delta}{2} \right)t \right]
\right.  \nonumber \\
& & \left. +\, \sum_{k=1}^{\infty} \frac{4\delta}{\pi k^2} \exp \left[- i
\frac{g}{\delta} \left( k + \frac{2\delta}{\pi k} \right) t \right] \right\}
,  \label{rho11}
\end{eqnarray}
where $\delta =gR/\pi c\ll 1$ is a dimensionless parameter,
characterizing the smallness of the cavity. These two extreme cases
are illustrated in Fig.~\ref{impu}, where we plot the degree of
impurity as a function of the time.

{\it Entanglement}\\
\indent Let us now examine the degree of entanglement of the states
described by the density matrix (\ref{rhored}) and how it evolves in
time. Let us initially calculate the concurrence~\cite{Wooters98}
associated with the density matrix $\rho$. The ``spin-flipped''
state is $\tilde{\rho}=(\sigma _2\otimes \sigma _2)\rho ^{*}(\sigma
_2\otimes \sigma _2)$, where $ \sigma _2=\left(
\begin{array}{cc}
0 & -i \\
i & 0
\end{array}
\right) $ is the Pauli matrix. Taking the basis
$\{|0_{+},0_{-}\rangle ,|0_{+},1_{-}\rangle ,|1_{+},0_{-}\rangle
,|1_{+},1_{-}\rangle \}$, we find
\begin{equation}
\rho \tilde{\rho}=\left(
\begin{array}{cccc}
0 & 0 & 0 & 0 \\
0 & \,2b(t)c(t)\, & 2b(t)d(t) & 0 \\
0 & 2c(t)d^{*}(t) & \,2b(t)c(t)\, & 0 \\
0 & 0 & 0 & 0
\end{array}
\right) ,
\end{equation}
where we have used that $b(t)c(t)=|d(t)|^2$. The concurrence of a
mixed bipartite state $\rho$ is given by~\cite{Wooters98}
\begin{equation}
C_\rho = {\rm max} \{ 0, \lambda_1 - \lambda_2 - \lambda_3 -
\lambda_4 \}
\end{equation}
where the $\lambda_i$s are the square roots of the eigenvalues of
the non-Hermitian matrix $\rho \tilde{\rho}$, written in decreasing
order. In our case, there is only one nonvanishing eigenvalue of
$\rho \tilde{\rho}$, given by $\lambda (t)=4b(t)c(t)$. We then find
the concurrence as
\begin{equation}
C_\rho (t) =2\sqrt{b(t)c(t)}  = 2 \sqrt{\xi (1-\xi )}\,|f_{++}(t)|.
\label{concur}
\end{equation}

\begin{figure}[th]
\includegraphics[{height=5.5cm,width=8.0cm}]{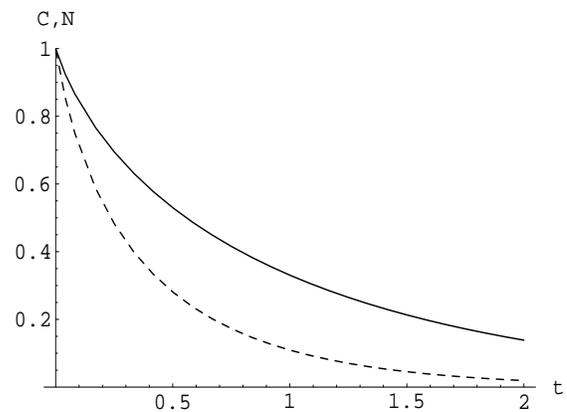}
\caption{Concurrence (full line) and negativity (dashed line) as
functions of time, for states with $\xi = 0.5$, ${\bar \omega} =
1.5$ and $g = 1.0$ (in arbitrary units), in a very large cavity.}
\label{CNg}
\end{figure}

\begin{figure}[th]
\includegraphics[{height=5.5cm,width=8.0cm}]{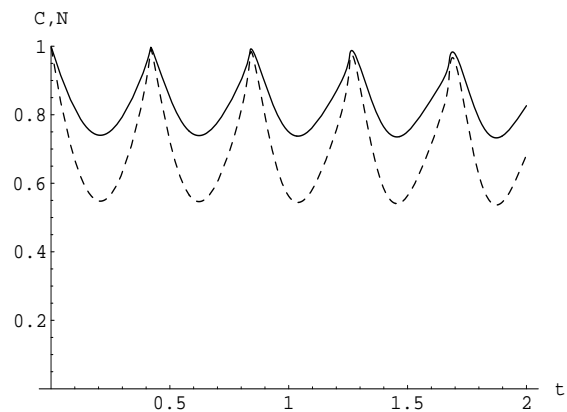}
\caption{Same as in Fig.~\ref{CNg}, but for a small cavity with
$\delta = 0.1$.} \label{CNp}
\end{figure}

Another measure of entanglement is the
negativity~\cite{VW02,andenaert}, which can be defined by
\begin{equation}
\mathcal{N}_\rho = \Vert \rho ^{T_{-}}\Vert_1 - 1
\end{equation}
where $\rho^{T_{-}}$ is the partial transpose of the bipartite mixed
state $ \rho $ and $\Vert \cdot \Vert _1$ denotes the trace norm.
The trace norm of an operator $\mathcal{O}$ is defined by $\Vert
\mathcal{O}\Vert _1=\mathrm{Tr}\sqrt{\mathcal{O}\mathcal{O}^{\dagger
}}$ which reduces, for hermitian operators, to the sum of the
absolute values of its eigenvalues. It can be easily shown that, for
hermitian operators,
\[
\Vert \mathcal{O}\Vert _1=\sum_j|\lambda _j|=\sum_{\lambda _j>0}\lambda
_j-\sum_{\lambda _j<0}\lambda _j=1+2\sum_{\lambda _j<0}|\lambda _j|,
\]
so that the negativity is given by the absolute value of the sum of the
negative eigenvalues of $\rho ^{T_{-}}$, that is,
\begin{equation}
\mathcal{N}_\rho = 2 \left| \sum_{\lambda _j<0}\lambda _j\right| .
\end{equation}
The partial transpose of $\rho$, $\rho^{T_{-}}$, is given by
\begin{equation}
\rho^{T_{-}}(t) = \left(
\begin{array}{cccc}
a(t) & 0 & 0 & d(t) \\
0 & b(t) & 0 & 0 \\
0 & 0 & c(t) & 0 \\
d^{*}(t) & 0 & 0 & 0
\end{array}
\right) ,  \label{rhotransp}
\end{equation}
whose eigenvalues are $\lambda_1(t) = b(t)$, $\lambda_2(t) = c(t)$,
$ \lambda_3(t) =(a(t) + \sqrt{a^2(t) + 4 |d(t)|^2})/2$ and
$\lambda_4(t) =(a(t) - \sqrt{a^2(t) + 4 |d(t)|^2})/2$. We thus
obtain the negativity as
\begin{eqnarray}
\mathcal{N}_{\rho}(t) & = & 2 |\lambda_4(t)| = \sqrt{a^2(t) + 4
|d(t)|^2} - a(t)  \nonumber \\
& = & \sqrt{\xi^2 + (4\xi - 6\xi^2)|f_{++}(t)|^2 + \xi^2
|f_{++}(t)|^4}  \nonumber \\
& &  -\, \xi + \xi |f_{++}(t)|^2 .
\end{eqnarray}

Notice that, for pure bipartite states, the concurrence and the
negativity, as defined above, are equal to each
other~\cite{andenaert}. However, for mixed bipartite states one has
$\mathcal{N}_{\rho} \leq C_{\rho}$~\cite{verstraete}. These features
are illustrated in Fig.~\ref{CNg} for a very large cavity and, in
Fig.~\ref{CNp}, for a small one, where we plot the concurrence and
the negativity for some of the states~(\ref{rhored}), as a function
of time.

{\it Conclusions}

Our study of how an entangled non-gaussian biatomic state contained
in a cavity evolves in time, leads to, as an overall conclusion,
that the behavior of the system is very contrasting in the cases of
a very large cavity (free space) or of a small cavity. This is
essentially due to the behavior of the quantity $|f_{++}(t)|^2$,
which governs the time evolution of the system, in each case. This
quantity (the probability that the center of mass oscillator remain
excited at the first level at time $t$), goes monotonically to
$zero$ as $t\rightarrow \infty$ for a very large cavity, while it
oscillates indefinitely with a strictly positive minimum for a small
cavity. These rather distinct behaviors are exhibited by the degree
of impurity as illustrated in Fig.~\ref{impu}.

With regard to entanglement, we have examined the concurrence and
the negativity as measures of entanglement; these are plotted in
Figs.~\ref{CNg} and ~\ref{CNp} for a very large and a small cavity,
respectively. In a very large cavity, these quantities decrease
monotonically to zero for large times and do not show any sudden
death. In contrast, for a small cavity, both present an oscillatory
behavior as time evolves and never vanish, having finite values for
any arbitrarily long elapsed time. This can be related to former
results in other context, obtained using dressed states, for decay
and stability of excited atoms in very large or small cavities, for
both zero or finite
temperature~\cite{termica1,termica,termalizacao,livro}.

Nevertheless it is worthy mentioning that, in some cases, a finite
asymptotic entanglement persists for long times in free space, that
is in very large cavities, as pointed out in
Refs.~\cite{prausner,benatti,liu,an,horhammer}. However, in most
cases, this phenomenon exists when the environment is a thermal bath
and for initial states that are two-mode squeezed (gaussian) states,
with a high degree of squeezing. These situations are different from
the case of the zero-temperature bath and non-gaussian initial
states we have considered in this note, for which no asymptotic
entanglement for large cavities exists.

In our case, for large cavities the entanglement vanishes
asymptotically. On the other hand, we have a {\it permanent},
oscillating  entanglement for small cavities, at all times. In the
case of a bath as we have adopted here, the existence of the
entanglement for all times in a sufficiently small cavity, in
contrast to the fast decay for long times in the very large cavity
(free space), favors small cavities as candidates to engender
phenomena that require entanglement as a resource. The discussion
presented here may be extended to other systems relevant to quantum
computation.

{\it Acknowledgments}

This work received financial support from CAPES, CNPq and FAPERJ,
Brazilian Agencies.


\begin{thebibliography}{99}
\bibitem{footnote} Distinctly from Refs.~\cite{pazes1,pazes2}, we do not
consider a direct coupling between the oscillators, but only their
interaction mediated by the field.
\end{thebibliography}
\end{document}